\documentclass[useAMS,usenatbib]{mn2e}
\usepackage{txfonts,graphicx}

\title[Type Ia supernovae and stellar winds in relativistic bubbles]
{Type Ia supernovae and stellar winds in AGN driven relativistic 
bubbles}
\author[N. N. Chugai et al.]
{N. N. Chugai$^{1,2}$\thanks{E-mail:nchugai@inasan.ru}, 
E. M. Churazov$^{1,3}$, R. A. Sunyaev$^{1,3}$\\
$^{1}$Max-Planck-Institut f\"ur Astrophysik,
   Karl-Schwarzschild-Str. 1, D-85741 Garching, Germany\\
$^{2}$Institute of Astronomy of Russian Academy of Sciences,
   Pyatnitskaya St. 48, 109017 Moscow, Russia\\
$^{3}$Space Research Institute, Profsoyuznaya 84/32, Moscow 117810, Russia\\
}
\begin{document}

\maketitle

\begin{abstract}
We analyse behavior of stellar winds of evolved stars 
and the outcome of SN Ia explosions in the AGN driven 
relativistic bubble. We find that the expansion of wind shells 
is efficiently decelerated by the relativistic pressure; their bulk 
motion however is preserved so they cross the bubble together 
with the parent star. The wind material occupies a small fraction of bubble 
volume and does not affect substantially the expansion of SN remnants. 
The estimated maximal radius of a SN remnant in the bubble is 30-40 pc,
if the envelope keeps its integrity and remains spherical. A fragmentation of 
SN shell due to Rayleigh-Taylor instability can alleviate the 
propagation of the SN material so the ejecta fragments 
are able to cross the relativistic bubble. Outside the bubble wind shells and 
supernova fragments are decelerated in the intracluster medium at 
close range off the bubble boundary. The deposited SNe~Ia material 
can enrich the intracluster gas with metals in a thin layer at the 
boundary of the relativistic bubble. This process may lead to a rim of 
enhanced line emission. 
In the opposite limit, when the fragmentation of supernova remnant is 
moderate or absent, the SN~Ia matter is advected by the relativistic 
plasma and may leave the central region of the bright cluster galaxy 
together with buoyantly moving bubbles.
\end{abstract}

\begin{keywords}
galaxies: clusters.
\end{keywords}

\section{Introduction}

Radio and X-ray observations show that in majority of regular galaxy
clusters, possessing a cool core, the activity of a central
supermassive black hole mediated by AGN jets
creates bubbles of relativistic plasma in the intra-cluster medium (ICM)
\citep{Boer93, HS98, Chur00, McN00, Bir04}. 
 X-ray data are consistent with the
assumption that bubbles are completely devoid of thermal gas 
\citep{Sand07} 
although the limits on the amount of thermal gas in the bubbles are
not very tight. The absence of strong shocks in the ICM surrounding the
bubbles implies that the relativistic plasma is in approximate
pressure equilibrium with the ICM.
These bubbles, inflated by an AGN, are believed to be responsible for
mechanical coupling of the AGN energy release and the thermal state of
the ICM in galaxy clusters, groups and individual elliptical
galaxies. For this reason every aspect of bubble physics receives much
attention. 

Given the ubiquity of bubbles in clusters, these objects
should be almost always present in the cluster core. 
The lifetime of the bubble in the stratified atmosphere is set by
``buoyancy'' time, while the growth rate of the bubble is defined by the
power of the AGN \citep[e.g.,][]{GN73, Chur00}. Pairs of bubbles are
located on both sides of the central supermassive black hole and often
several generation of bubbles differing in their size and the distance
from the center are observed. 

Bubbles with sizes of order 1-10 kpc are often found in the inner regions of 
galaxy clusters, sharing the space with brightest cluster galaxy 
(BCG). 
For example, in NGC 1275 (BCG of Perseus cluster) within the 
effective radius of the galaxy ($R_e\sim$15 kpc) we see two bubbles on 
either side of the nucleus with the radius of each bubble $\sim6.3$ kpc 
\citep[e.g.,][]{Boer93, Fab03}. In  M 87 (Virgo cluster) 
the bubble radius is $\sim1.4$ kpc \citep[e.g.,][]{For05, For07}, while
the galaxy effective radius is of order 7.7 kpc. We expect that in both cases
bubbles should contain significant fraction of the galaxy stars. These
old low mass stars evolve as usual, lose their mass via the wind, and 
some of them give birth to type Ia supernovae (SNe\,Ia). 
Even if the relativistic bubble sweeps up all the
thermal gas in the process of inflation a lot of material in the
form of the stellar wind thus can be supplied by the stars {\it inside
the bubble} during the bubble life. This wind material could affect
the expansion dynamics of supernovae exploded in the bubble.

The question we address here is what happens to the stellar winds
and SNIa ejecta embedded in the relativistic bubble. Almost weightless
relativistic plasma provides highly unusual conditions for the dynamic
evolution of wind shells and SNe. Two extreme scenarios are conceivable.
In one limit all the wind material and SN ejecta are decelerated and 
mixed with relativistic plasma. During subsequent evolution the
buoyantly rising bubbles advect this material with them. In another
limit the matter ejected by evolving stars propagates freely
through the relativistic plasma, attains the boundaries of the bubble
and enriches the ICM just outside the bubble with heavy elements.
We investigate different scenarios of dynamical evolution of wind shells 
and SN ejecta and explore observational outcomes of these scenarios.

The structure of the
paper is as follows. We first study the expansion and bulk motion of
the wind envelopes including mass striping effects. This provides us 
with the estimate of the filling factor of the wind shells 
ensemble. We then address the issue of the SN envelope expansion dynamics
and bulk motion, 
Rayleigh-Taylor fragmentation of the SN shell and propagation of ejecta 
fragments in the relativistic plasma. Finally, we discuss the implications 
of results for the relativistic bubble matter contents and intracluster 
thermal environment. 
 
\section{Wind material in relativistic bubble}
\label{sec-wind}

\subsection{Stars and mass injection rate}
\label{sec-stars}

We consider, as a fiducial model, a spherical bubble of 
radius $R_b=5$ kpc with the bubble center at the distance $R_b$
from the center of the BCG.  The characteristic age of this bubble is
$t_b\sim2R_b/v_b\sim3.3\times10^7$ yr, where we adopt the bubble rise
velocity $v_b=300$ km s$^{-1}$ \citep[cf.][]{Chur01}.  At the
radii $r<15$ kpc the stellar component in massive elliptical galaxies
dominates \citep[e.g.'][]{John09} and its density distribution 
can be approximated by the singular
isothermal sphere with the velocity dispersion $\sigma_v$
\begin{equation}
\rho=\frac{\sigma_v^2}{2\pi Gr^2}=\rho_0\left(\frac{r_0}{r}\right)^2\,,
\end{equation}
where $G$ is gravitational constant and $r_0$ is a radial
scale\footnote{This approximation breaks at large radii to ensure
  convergence of the total stellar mass.}. 
Adopting $r_0=10$ kpc and velocity dispersion  
$\sigma_v=350$ km s$^{-1}$ \citep{WT06} one gets 
the stellar mass  
$M(<r_0)=M_0=5.6\times10^{11}~M_{\odot}$. This value is consistent 
with estimates 
of the total mass of stellar component of BCG of $\sim10^{12}~M_{\odot}$.
Inside the bubble of radius $R_b=5$ kpc the stellar mass in this case is 
$M_s\approx9\times10^{10}R_{b,5}~M_{\odot}$, where $R_{b,5}=R_b$/(5 kpc). 
 
The stars are presumably old with an age of $t\sim 10^{10}$ yr, which suggests
that the current upper limit of the stellar mass at the AGB stage is $\approx
1~M_{\odot}$ \citep{Scha92}. Assuming 
the Salpeter initial mass
function $dN/dm=Cm^{-\alpha}$ within the range $m_1<m<m_2$ one gets
the normalizing factor
\begin{equation}
C=(\alpha-2)M_s[m_1^{-(\alpha-2)}-m_2^{-(\alpha-2)}]^{-1}=0.28M_s\,,
\label{eq-normf}
\end{equation}
where $\alpha=2.35$, $m_1=0.1~M_{\odot}$, and $m_2=1~M_{\odot}$ 
are used.

The star of $m=1~M_{\odot}$ leaves behind a white dwarf of 
$m_{wd}=0.5~M_{\odot}$ \citep{Sala09}, while 
$\Delta m=m-m_{wd}=0.5~M_{\odot}$ is lost in the form of slow 
($u\approx10-30$ km s$^{-1}$) wind during the thermally pulsing AGB stage  
\citep{VW93}. The present day 
integrated rate of the wind matter injection into the relativistic bubble is 
\begin{equation}
\dot{M}=(m_2-m_{wd})\frac{dN}{dm}\frac{dm_2}{dt}\,,
\label{eq-dotmw}
\end{equation}
where $dm_2/dt$ is the rate at which upper limit $m_2$ 
decreases with time. This rate is determined by the relation between 
the lifetime and initial star mass $t=t_2(m_2/m)^{\beta}$, where $\beta=3.2$ 
in the range of 
$1-2~M_{\odot}$ \citep{Scha92}. For these values and 
$M_s=9\times10^{10}R_{b,5}~M_{\odot}$ the equation (\ref{eq-dotmw}) yields 
$\dot{M}=0.416~M_{\odot}R_{b,5}$ yr$^{-1}$. The corresponding stellar 
death rate is $\dot{N}=0.83R_{b,5}$ yr$^{-1}$ which is also 
the rate of wind shell formation $\dot{N_w}$. 
One expect thus to find $N_w=t_b\dot{N_w}\approx2.7\times10^7R_{b,5}$ 
newly created wind envelopes in the bubble volume with  
the total amount of the wind matter in the bubble of  
$0.5N_w\approx 1.4\times10^{7}R_{b,5}~M_{\odot}$, factor $\sim10^4$ larger 
than the mass in the form of relativistic 
particles of the bubble, $3pV/c^2\sim2.4\times10^3~M_{\odot}$. 
It should be noted, however, that the $N_w$ estimate ignores so far 
a possible escape of wind envelopes from the bubble, which is addressed 
below. 

\subsection{Wind shell dynamics}
\label{sec-windyn}

The dynamical effect of the  wind matter in the bubble on a certain
wind shell or SN ejecta 
depends on the filling and covering factors of wind shells. To assess the 
situation one needs first to find 
the average volume and size of a wind envelope at the final stage of its 
expansion. 
Here we assume that the bubble is static and adopt that a star moves 
with the mean velocity $v_s$. Despite a singular isothermal 
sphere is assumed for the stellar population, it is reasonable to 
estimate $v_s$ 
using Maxwell velocity distribution truncated at the escape velocity 
$v_e$. The truncated Maxwell distribution is taken in the form 
proposed by \citep{King66}: 
$f(v)\propto[\exp(-v^2/2\sigma_v^2) - \exp(-v_e^2/2\sigma_v^2)]$
 for $v<v_e$ and $f=0$ otherwise.
Adopting $\sigma_v=350$ km s$^{-1}$, i.e., $v_e=2\sigma=700$ km s$^{-1}$ 
we come to the average velocity $v_s=400$ km s$^{-1}$. 

The geometry of the wind shell that forms as a result of 
the interaction of the wind with the relativistic medium
depends on the value of the drag force exerted on the wind boundary. 
If the drag force 
is strong the stripped wind material creates a trailing plume. On the 
other hand, if drag force is very weak, then the wind shell moves with 
the star velocity retaining spherically-symmetric shape and eventually 
may escape the relativistic bubble. To explore this issue we consider 
major stages of the mass loss at the AGB and post-AGB stage: 
\citep{Stef98, LZ10}:
(1) the slow wind ($u=10$ km s$^{-1}$, $\dot{M}\sim 10^{-7}~M_{\odot}$ yr$^{-1}$)
on the time scale of the AGB stage, i.e., $10^6$ yr, (2) slow superwind 
($\dot{M}\sim 10^{-5}-10^{-4}~M_{\odot}$ yr$^{-1}$)
during the last $\sim10^4$ yr of
the AGB stage, and (3) fast wind at the post-AGB stage  
which corresponds to the planetary nebula (PN) stage  ($\sim10^4$ yr). 
The last stage practically does not contribute to the mass loss,  
but turns out essential for the acceleration of the slow wind.
We adopt that 60\% of the hydrogen shell is lost at 
the first stage and 40\% at the superwind stage 
\citep{Stef98}, which implies  
the mass-loss rates 
$\dot{M}\sim 3\times10^{-7}~M_{\odot}$ yr$^{-1}$ at the slow wind stage and 
$\dot{M}\sim 2\times10^{-5}~M_{\odot}$ yr$^{-1}$ at the superwind stage.
For the fast wind stage  we adopt parameters derived from 
the modelling of the X-ray emission of the PN:  
$\dot{M}\sim 2\times10^{-8}~M_{\odot}$ yr$^{-1}$ and $u=1500$ km s$^{-1}$ 
\citep{LZ10}. 
The fast wind parameters suggest that the total kinetic energy 
released during this stage ($\sim10^4$ yr) is $E_3\approx4.5\times10^{45}$ erg. 
When transfered to the slow wind shell with the mass 
of $0.5~M_{\odot}$ this energy accelerates the shell up to 
$u\approx30$ km s$^{-1}$,  
in accord with the expansion velocities of evolved PN \citep{Rich08}.

The mass striping rate of the wind shell scales as the shell radius 
squared, so at the stage of slow wind the maximal 
stripping is attained at the end of this stage. The radius of the 
wind shell at this age can be estimated from the energy arguments.
The kinetic energy of the wind shell is spent on the 
$pV$ work against external pressure $p$ and on the internal energy 
which results in the stopping radius of the wind shell 
\begin{equation}
r_1=\left[\frac{3}{8\pi}\left(\frac{\gamma-1}{\gamma}\right)
\frac{M_1u_1^2}{p}\right]^{1/3}
= 0.19p_{10}^{-1/3}~~\mbox{pc}\,,
\label{eq-rwind}
\end{equation}
where $p_{10}=p/(10^{-10}~\mbox{erg cm}^{-3})$, 
$\gamma=5/3$, $u_1=10$ km s$^{-1}$, and $M_1=0.3~M_{\odot}$ are used.

The wind shell moves as a whole together with the 
white dwarf unless the bulk of the material is stripped into the 
trailing plume. 
The wind shell stripping can be estimated following \citet{Nu82} 
consideration of the gas stipping for a galaxy moving in the ICM. 
The turbulent stripping is determined by the combined effect of 
the Kelvin-Helmholtz instability (KHI) and the ram pressure drag. 
Indeed, KHI broadens the boundary layer which results in the 
supression of the KHI. The net stripping rate, therefore, is 
controlled by the ram pressure
\begin{equation}
\dot{M}=\pi r_1^2\rho_a v_s=
1.6\times10^{-12}p_{10}^{1/3}~~M_{\odot}~\mbox{yr}^{-1}\,,
\label{eq-dotm}
\end{equation}
where we use $\rho_a=3p/c^2$ for the density of ambient medium.
The average residence time of a wind shell in the bubble is 
 $R_b/v_s\sim1.2\times10^7$~yr, so the above mass loss rate implies that 
the wind shell loses $\sim2\times10^{-5}~M_{\odot}$ while moving in the 
bubble, negligibly small amount compared to the mass of the wind shell, 
$0.3~M_{\odot}$. 

Alternatively, the stripping could be caused by the  Alfven wave drag. 
In this regard we note that the infinite conductivity approximation 
for the wind shell is fully applicable.  
A conducting body moving with velocity $v$ across the magnetic field $B$ 
experiences the drag force due to Alfven wave generation 
\citep{Dre65} 
\begin{equation}
F_d=(B^2/4\pi)(v/v_A)S\,,
\label{eq-dforce}
\end{equation}
where $S$ is the area of lateral surface perpendicular to $\vec{B}$,
$v_A$ is the Alfven velocity, $v_A\approx B/\sqrt{4\pi \rho_a}$ with 
$\rho_a=3p/c^2$. Strictly speaking, the Alfven velocity in the 
relativisic plasma \citep{Ged93} is smaller compared to this 
expression by a factor of 0.7-0.9 depending on 
the ratio of magnetic to total pressure; we neglect this difference. 
For a sphere of the radius $r$ the lateral area is
$S\approx 2\pi r^2$ and the mass stripping rate caused by the Alfven 
wave drag is  
\begin{equation}
\dot{M}=\frac{B^2r_w^2}{2v_A}\approx4.1\times10^{-10}p_{10}^{7/6}
B_5~M_{\odot}~\mbox{yr}^{-1}\,,
\label{eq-dotma}
\end{equation}
where $B_5=B/(10^{-5}~\mbox{G})$.
The stripping rate due to the Alfven drag thus turns out two orders of 
magnitude larger than 
the rate according to equation (\ref{eq-dotm}). Yet even for the 
Alfven drag the mass lost during the residence time 
is only $\sim0.01~M_{\odot}$ which is a small fraction ($\sim3$\%) 
of the wind shell. We thus conclude that the wind shell lost at the 
slow wind stage 
remains almost intact while traveling across the bubble.
A similar result can be obtained for the superwind stage.
The outcome of a combined effect of all three stages of the mass loss 
is a spherical wind shell of $0.5~M_{\odot}$ expanding 
with the velocity of 30 km s$^{-1}$. Using equation (\ref{eq-rwind})
one finds the wind shell stopping radius is $r_w=0.5p_{10}^{-1/3}$ pc.

The above treatment of the wind dynamics suggests that the 
cosmic rays diffusion into the shell can be neglected. To check whether 
this assumption is valid we assume Bohm diffusion coefficient $r_gc/3$, 
where $r_g$ is the proton giroradius. 
This assumption is standard for the analysis of cosmic ray propagation 
and supported by observational data on the cosmic ray acceleration in 
supernova remnants \citep{Stage06}, although the concept of 
a tangled field might seriously modify a picture 
of the cosmic ray diffusion in magnetic field \citep{Nar01}.
For the spectral index of relativistic protons $>2$ the 
energy of cosmic rays resides in low energy protons, which permits us to use  
the characteristic energy of relativistic protons $\sim 1$ GeV.
With $B=10^{-5}$ G  one gets $r_g\sim3\times10^{11}$ cm. 
The diffusion time is then 
\begin{equation}
t_{d}\sim\frac{r_w^2}{r_gc}=3\times10^6B_5^{-1}r^2_{w,18}~~\mbox{yr}\,,
\end{equation}
where $r_{w,18}=r_w/(10^{18}\,\mbox{cm})$. For the slow wind 
stage with the final radius of $r_1\sim0.2$ pc the 
diffusion time is comparable to the duration of this stage ($\sim10^6$ yr), 
while the life time of superwind and fast wind stages
is significantly smaller than the diffusion time.

The estimated time scale of the cosmic ray diffusion 
suggests that the penetration of cosmic rays in the wind 
can affect the wind expansion dynamics at the slow wind stage making the 
pressure gradient smoother and the deceleration less pronounced. As a 
result, the final radius of 
the wind shell at the slow wind stage in fact could be somewhat larger, 
$r_1>0.2$ pc. On the other hand, the effect cannot be significant 
because the diffution time increases $\propto r_1^2$, so the 
role of the cosmic ray diffusion rapidly drops for larger radius. 
We conclude therefore that the stopping 
radius of the wind shell ($r_w\sim0.5$ pc), which includes 
the combined effect of AGB and post AGB mass loss and 
omits the cosmic ray diffusion, is a reasonable estimate. 

\subsection{Wind shell escape}
\label{sec-winesc}

Outside the relativistic bubble the wind shell turns out in 
the intracluster thermal gas. For the mass stripping rate 
$\dot{M}=\pi r_w^2\rho_av_s$ adopting the wind shell radius 
$r_w=1.5\times10^{18}$ cm, number
 density of interstellar gas $n=0.02$ cm$^{-3}$, and $v_s=400$ km s$^{-1}$ 
 one obtains 
$\dot{M}\sim1.7\times10^{-7}~M_{\odot}$ yr$^{-1}$. 
It takes $3\times10^6$ yr to completely strip 
the $0.5~M_{\odot}$ wind shell over the distance of $\sim1$ kpc. 
The same estimate can be obtained from the equation of deceleration 
of the wind shell as a whole by drag force $\pi r_w^2\rho_av_s^2$. 
The wind shell escaping bubble is decelerated thus in a close vicinity 
of the bubble boundary. 

The average residence time of the wind shell in the bubble 
$R_b/v_s\sim1.2\times10^{7}$ yr is somewhat smaller than the age of 
the fiducial 
bubble $3\times10^7$ yr. The total amount of the wind shells residing 
in the bubble is, therefore, $\dot{N}R_b/v_s\sim10^7$, while 
the filling factor of the ensemble of wind shells in the bubble is 
\begin{equation}
f=N_w(R_w/R_b)^3\sim10^{-5}p^{-1}_{10}\,.
\label{eq-ffac} 
\end{equation}
The probability of a collision with the wind shell is determined by the 
ratio of the bubble radius and the mean free path. The latter is  
\begin{equation}
\lambda=(\pi r_w^2n_w)^{-1}=64p^{2/3}_{10}~~\mbox{kpc}
\label{eq-taug} 
\end{equation}
where $n_w=(3/4\pi)N_w/R_b^3=2\times10^{-5}$ pc$^{-3}$ 
is the number density of wind shells. The probability 
of shell collisions is low, because the average number of wind shells 
along the bubble radius is only $\tau=R_b/\lambda=0.08$.
The average probability of the collision is approximately 
$\approx[1-\exp(-\tau)]\approx\tau=0.08$. 
More accurate estimate can be done using expression for 
the escape probability of the 
photon from a homogeneous sphere \citep{Ost89}
\begin{equation}
p_{\rm esc}=\frac{3}{4\tau}\left[1-\frac{1}{2\tau^2}+\left(\frac{1}{\tau}+
\frac{1}{2\tau^2}\right)
\exp (-2\tau)\right]\,,
\label{eq-oster}
\end{equation}
For $\tau=0.08$ the 
equation (\ref{eq-oster}) gives $p_{\rm esc}=0.94$. Most of wind shells 
therefore escape the bubble freely and only 6\% of 
wind shells collide with another shell. 

The fate of the collided wind shells depends on whether the collision 
is adiabatic or radiative. With the average relative velocity of 
collision $u\sim560$ km s$^{-1}$ and the wind shell density 
$\sim200$ cm$^{-3}$ the estimated cooling 
time of the shocked gas turnes out to be  $\sim3\times10^{10}$ s, comparable 
with the hydrodynamic scale $r_w/u\sim3\times10^{10}$ s. This means 
that both adiabatic and radiative collision regimes are plausible. 
In the adiabatic case wind shells approximately retain their sizes and absolute 
velocities, so the adiabatic collision 
does not affect their escape. In radiative case 
collided shells merge and form thin dense pancake of a thickness $b\ll r_w$  
and density $\rho_c\sim(r_w/b)\rho_w$, 
where $\rho_w$ is the density of the wind shell before collision,
This pancake is liable to fragmentation into clumps of size $a\gtrsim b$ 
and average velocity $\sim v_s/\sqrt{2}$. It is easy to verify that for 
fragments with
sizes  $a\gtrsim b$ and density $\rho_c\sim(r_w/b)\rho_w$
the stripping (or deceleration) time is greater 
than the stripping time of wind shells. 
We thus conclude that the most of the wind material 
escapes into the hot ICM.

In our analysis of the wind shell dynamics we ignored a possible fragmentation 
of the wind shell due to the Rayleigh-Taylor (RT) instability on the 
deceleration or acceleration stages. This omission facilitates the 
consideration; yet it does not affect the major conclusion that the wind 
shell material escape the relativistic bubble. As we will see below, 
the RT fragmentation favours the shell matter escape. 

\section{Type Ia supernovae in relativistic bubble}

With SN\,Ia production efficiency $\psi=0.008$ per one white dwarf 
formed in the stellar population of 
E-galaxies \citep{Prit08} and the stellar death 
rate in the bubble of fiducial model $\dot{N}=0.83$ yr$^{-1}$ 
one expects $\psi\dot{N}t_b\sim2\times10^5$ SN\,Ia explosions 
 during the 
relativistic bubble lifetime $t_b=3\times10^7$ yr. We now consider in 
detail SN expansion in the relativistic bubble and analyse an outcome 
of the Rayleigh-Taylor fragmentation of decelerating SN shell.

\begin{figure}
\includegraphics[width=85mm]{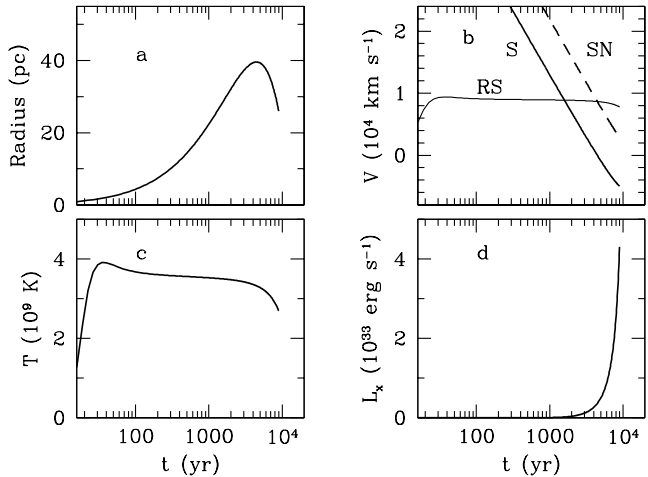}
\caption{
Thin shell model of supernova evolution in relativistic bubble.
Shown are the shell radius ({\bf a}), evolution of velocities 
of the shell (S), pre-shock velocity of supernova ejecta (SN), 
and reverse shock speed (RS) ({\bf b}), 
evolution of the reverse shock temperature ({\bf c}), 
X-ray luminosity ({\bf d}).
   }
\label{f-tshell}
\end{figure}

\subsection{SN expansion}

Given a small number of wind shells along the bubble radius
(Section \ref{sec-winesc}) the deceleration of the SN expansion 
in the relativistic bubble is probably dominated 
by the pressure of the relativistic fluid. Indeed, 
for SN expanding with the characteristic velocity 
$v\approx(2E/M)^{1/2}=10^9$ cm s$^{-1}$ 
one readily sees that the ram pressure is small compared to 
the pressure of relativistic fluid: $\rho v^2=3p(v/c)^2\ll p$. 
The crucial role of the 
external relativistic pressure in the SN deceleration is a distinguishing 
feature compared to the standard case of SN shell in the ordinary
interstellar medium. 

In our analysis of the SN expansion we assume the isotropic pressure of the 
relativistic medium. This is the case, if the mean free path for 
relativistic protons along the magnetic field is much less than the SN radius. 
Since the SN expands subsonically relative 
to the external medium, in which the sound speed is $\approx c/\sqrt3$, the 
strong forward shock does not form. 
The reverse shock obviously forms, because outer layers of ejecta are 
decelerated by the external pressure and the velocity jump between 
the undisturbed ejecta and swept-up shell exceeds the sound speed 
in the unshocked ejecta ($\sim10$ km s$^{-1}$). The SN is fully 
decelerated when the 
reverse shock crosses the bulk of the ejecta mass. 

To estimate the stopping radius 
of SN   one can use the energy considerations 
likewise we did 
for the wind shell expansion. The initial kinetic energy $E$ of SN  should be 
spent 
on the $PV$ work against the external pressure and on the internal energy 
$pV/(\gamma-1)$ which gives the stopping radius
\begin{equation}
r_{sn}=\left[\left(\frac{\gamma-1}{\gamma}\right)\frac{3E}{4\pi p}
\right]^{1/3}=36p_{10}^{-1/3}~~\mbox{pc}\,.
\label{eq-rstop}
\end{equation}
 With a characteristic ejecta velocity $v\approx10^9$ km s$^{-1}$ it takes 
$t_s\sim r_{sn}/v\sim3\times10^3$ yr to reach $r_{sn}$, rather short 
time compared to the bubble lifetime.  

The dynamics of the swept-up shell and the X-ray emission of the 
reverse shock can be illustrated using a model based on 
a thin shell approximation. 
This suggests that the shell formed by ejecta material flowing 
into the reverse shock is considered as a thin shell which dynamics 
is governed by the dynamical pressure of the SN ejecta and external 
pressure $p=10^{-10}$ erg cm$^{-3}$. The equation of motion for the 
thin shell is
\begin{equation}
M\frac{dv}{dt}=4\pi r^2\left[\rho\left(\frac{r}{t}-v\right)^2-p\right]\,,
\label{eq-motion}
\end{equation}
where the shell mass is determined by the mass conservation 
\begin{equation}
\frac{dM}{dt}=4\pi r^2\rho\left(\frac{r}{t}-v\right)\,.
\label{eq-mass}
\end{equation}
These equations are solved numerically assuming a freely expanding 
SN with the mass 
of $1.4~M_{\odot}$, energy of $1.5\times10^{51}$ erg, the density distribution 
$\rho\propto \exp{(-v/v_0)}$, boundary velocity of 
$4\times10^4$ km s$^{-1}$, and initial outer radius of $10^{18}$ cm. 

Results are displayed in Fig. \ref{f-tshell} which 
shows the evolution of the shell
radius, velocity of the shell, boundary velocity of SN ejecta and velocity of 
the reverse shock, reverse shock temperature assuming full equilibration,  
and X-ray luminosity of the reverse shock. The maximal radius of the thin 
shell model is 39 pc, slightly larger than analytical estimate 
$r_{sn}=36$ pc. Remarkably, the thin shell shows 
contraction phase (Fig. 1a) which is a direct outcome of the 
dynamical role of the external pressure. However, since we neglect 
the internal pressure of the shocked envelope, the amplitude of the 
contraction phase 
in our model is exaggerated, so we stop the computations at this phase. 
To calculate the X-ray emission we assume 
that the hot plasma in the shell is distributed homogeneously in the shell 
with the thickness $\Delta R/R=0.1$. It is rather a crude approximation 
because the density distribution in the reverse shock is expected to be 
essentially inhomogeneous with a peak at the contact surface.
Yet it is reasonable enough to get an idea about X-ray luminosity 
within a factor of two. The luminosity is maximal at the 
contraction phase and reaches $\sim4\times10^{33}$ erg s$^{-1}$ 
at the shock temperature of $\sim200$ keV. The equilibration of electrons 
and ions, however, is an oversimplification, so the shock electron 
temperature and the luminosity should be considered approximate.

\begin{figure}
\includegraphics[width=80mm]{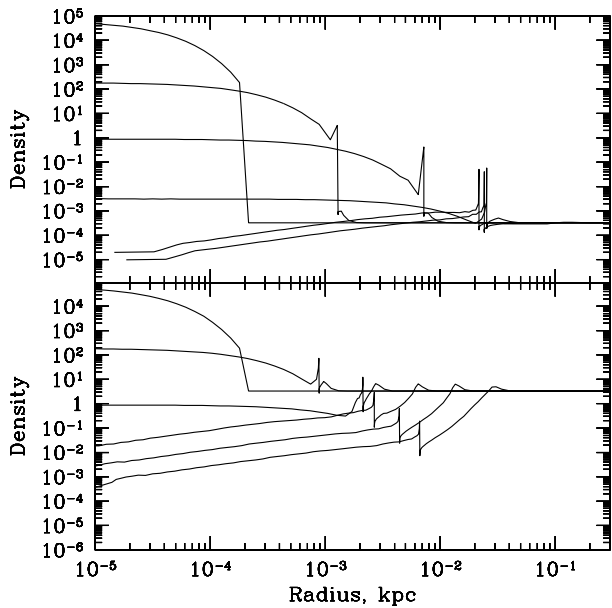}
\caption{
Density distribution for SNIa envelope expanding in the ICM at the ages
$\sim$60, 400, 3000, $2~10^4$ and $10^5$ years for 
the ICM temperature of 0.01 keV ({\bf bottom}) and 
 of 100 keV ({\bf top}). 
Position of a sharp wiggle in the density distribution corresponds 
to the contact discontinuity separating SN ejecta and the ICM. 
\label{f-snv}}
\end{figure}

Another interesting view on the SN expansion dynamics 
in the relativistic bubble 
gives us one-dimensional hydrodynamic simulations in which we 
assume a hot rarefied non-relativistic plasma to be a proxy 
for the relativistic fluid. The thermal pressure is taken the same 
$p=10^{-10}$ erg cm$^{-3}$.
The ICM temperature varies in different runs from
$10^{-2}$ up to $10^{4}$ keV\,\footnote{We use nonrelativistic equation
  of state in these illustrative runs even for $T_e=10^4$ keV,
  although this not valid for electrons.}. For $T=10^{4}$ keV the situation 
is close to the case of the relativistic medium because the 
thermal pressure exceeds the dynamical pressure in the upstream flow of 
the forward shock. We assume homologous
expansion of the envelope $v\propto r$ and model the initial density
distribution 10 years after the explosion as $\rho\propto e^{-r/r_0}$,
where $r_0=3~10^{-5}$ pc. The ejecta mass is $1.4~M_{\odot}$ and kinetic 
energy is $1.5\times10^{51}$ erg, while the maximum expansion velocity 
is set to $2~10^{4}~{\rm km~s^{-1}}$.

\begin{figure}
\includegraphics[width=80mm]{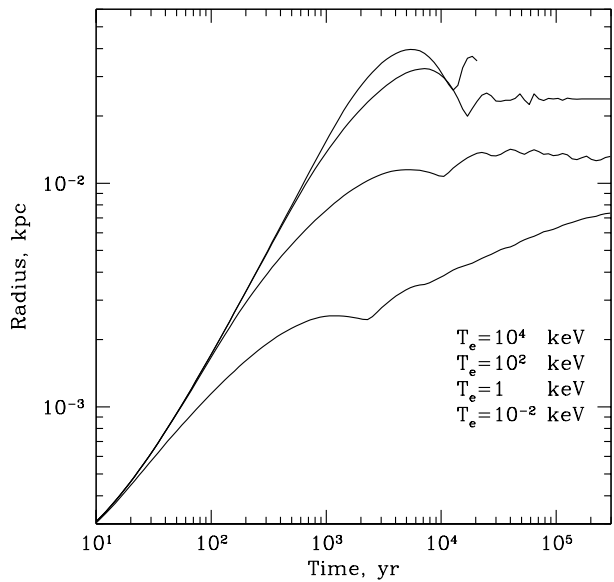}
\caption{Radius of the contact discontinuity, separating SN ejecta and
  the ICM as a function of time. Four curves shown correspond to
  explosions in the ICM with the same pressure, but different
  temperatures, 0.01, 1, 100, $10^4$ keV, from bottom to top. 
  Clearly the final size of
  the ejecta is largest in the ICM with highest temperature.
   }
\label{f-rb}
\end{figure}

\begin{figure}
\includegraphics[width=80mm]{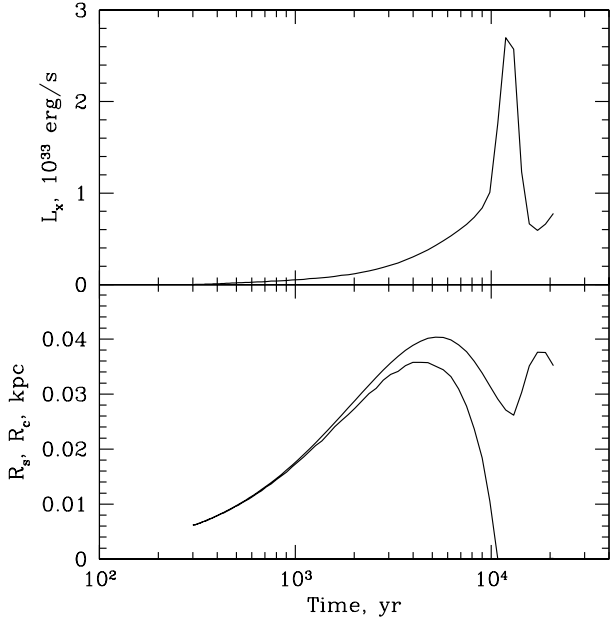}
\caption{X-ray luminosity of the reverse shock in the model 
with the ICM temperature of $10^4$ keV ({\bf top}) and the radii of 
the contact discontinuity and reverse shock ({\bf bottom}). 
The luminosity decreases after $\sim10^4$ yr because of 
the significant expansion of the postshock layer between 
the reverse shock ({\bf bottom}, lower curve) and the contact 
surface (upper curve).
   }
\label{f-xlum}
\end{figure}

The dependence of the expansion dynamics on the temperature of the ICM
(at the same pressure) is apparent from Fig.~\ref{f-snv}. In the low
temperature case (bottom panel in Fig.~ \ref{f-snv}) most of
ejecta energy is spent on a forward shock, which is barely
resolved in our simple simulations. By contrast, for the
high temperature ICM almost all the initial kinetic energy is eventually
converted into the enthalpy of the ejecta and only tiny amount of 
energy is deposited in the forward shock. Accordingly the final size of the
envelope at the boundary separating ejecta and ICM is much larger for
the high temperature run. This is further illustrated in
Fig.~\ref{f-rb}, showing the time dependence of the envelope
radius. The simulations show that in the 
limit of very hot ICM the final envelope radius converges to the value 
given by the thin shell model (Fig. \ref{f-tshell}).
The X-ray luminosity of the reverse shock in the model with 
$10^4$ keV medium is shown in Fig. ~\ref{f-xlum} together with the 
evolution of the radii of the reverse shock and contact surface. 
The luminosity evolution is consistent with the prediction of 
the thin shell model (Fig. \ref{f-tshell}) at the ages 
$\lesssim6\times10^{3}$ yr. However, at the final stage
of the ejecta deceleration 
($t\gtrsim10^4$ yr) the luminosity behavior differs from 
that of the thin shell model. Indeed at this phase the shocked gas 
cannot be treated as thin shell. 
After about $10^4$ yr the reverse shock attains the center. This is 
accompanied by the overall contraction; as a result the 
emission measure increases and the luminosity attains maximal value 
$\sim3\times10^{33}$ erg s$^{-1}$. 
This is followed by the expansion which results in the luminosity drop.
At the most luminous phase, $L_{\rm x}\gtrsim10^{33}$ erg s$^{-1}$, 
the SN~Ia remnant lives 
$\sim4\times10^3$ yr. The temperature of the shocked ejecta at this phase 
is in the range of $10^8-10^9$~K, so only a small fraction of the total 
luminosity ($10-30$\%) falls into the standard {\em Chandra} band 
(0.2-10 keV).

Generally, the SN expansion dynamics can be affected by the diffusion of
relativistic protons into the ejecta. This process might modify dynamics 
by smoothing out the pressure jump at the boundary between ejecta and 
relativistic fluid. 
The time it takes to fill the SN by cosmic rays with the energy $E$ per
particle can be estimated as the time for the proton to escape from the 
relativistic bubble layer adjacent to SN. The volume comparable with SN of 
radius $R$ 
is a spherical layer of a thickness of $\sim0.3R$. Assuming Bohm diffusion 
coefficient $D=cr_g/3$ one gets the diffusion time 
\begin{equation}
t_{d}\sim\frac{(0.3R)^2}{4D}=1.5\times10^9B_5
\left(\frac{R}{30\,\mbox{pc}}\right)^2~\mbox{yr}\,,
\end{equation}
The time it takes to fill the SN by cosmic rays at the 
essential deceleration epoch turns out tremendous compared to the SN age  
($\sim3\times10^3$ yr). We conclude, therefore, that 
the diffusion penetration of relativistic protons into the SN envelope
unlikely affects the SN expansion dynamics. 

\subsection{SN bulk motion}

After the expansion braking the SN shell still 
retains the bulk motion with the typical velocity 
$v_s=400$ km s$^{-1}$. If the deceleration of the 
bulk motion were negligible, the SN shell would escape the 
relativistic bubble after the average residence 
time $R_b/v_s\sim10^7$ yr. We now check, whether 
the ram pressure and the Alfven wave drag can substantially 
decelerate the bulk motion inside the relativistic bubble. 

The ram pressure drag force exerted on the SN shell is 
 $F_d=\pi r_{sn}^2\rho_av^2$,
where the ambient density is $\rho_a=3p/c^2$, assuming particles 
dominate in the pressure. Using the equation of motion 
\begin{equation}
M\frac{dv}{dt}=-\pi r_{sn}^2\rho_av^2\,, 
\label{eq-snram}
\end{equation}
the characteristic deceleration length can be estimated as 
\begin{equation}
l_d\approx \frac{Mc^2}{3\pi r_{sn}^2p}\approx 69p_{10}^{-1/3}~~\mbox{kpc}\,.
\label{eq-lramd}
\end{equation}
This shows that the deceleration of the bulk motion of SN by the ram 
pressure can be neglected. 

The Alfven wave drag exerted on the spherical SN shell is defined 
similarly to the case of the wind shell, i.e., 
$F_d=(1/2)B^2(v/v_A)r_{sn}^2$. The chracteristic deceleration  time 
is $t_d\approx v_sM/F_d$, while the deceleration length,  
$l_d\approx v_st_d$, is 
\begin{equation}
l_d\approx \frac{Mv_sc}{Br_{sn}^2\sqrt{12\pi p}}\approx 
0.3v_{s,400}B_5^{-1}p_{10}^{1/6}~~\mbox{kpc}\,,
\label{eq-ldec}
\end{equation}
where $v_{s,400}=v_s/(400~\mbox{km s}^{-1})$. This shows that 
the Alfven wave drag 
eesentially brakes the bulk motion of the SN shell at the distance 
much smaller than the radius of the relativistic bubble even for weak field 
$B\sim3\times10^{-6}$ G. 
We thus conclude that the ejecta of SN~Ia exploded in the 
relativistic bubble cannot escape the bubble due to the bulk motion.
Amazingly, the SN material is decelerated in the slow bulk motion
at the distance ten times larger than in the maximum radius attained in the high 
speed envelope expansion.

\subsection{Rayleigh-Taylor instability and spike deceleration}

The swept-up SN shell decelerating in the light relativistic 
fluid is liable to the Rayleigh-Taylor instability 
(RTI) which generally should result in the fragmentation of 
SN shell, close to the stage of the significan deceleration, i.e., 
at about the stopping radius $r_{sn}$. 
The situation is similar, albeit inverted with respect to Crab nebular. 
The Crab shell {\em accelerated} by the shocked relativistic wind
shows long thin RT spikes directed backward the center 
\citep{Hester96}. In case of decelerated SN 
dense RT spikes protruded forward could travel large distances before 
they stop. Yet it should be emphasised the difference with the Crab. 
In the latter case the SN material pressurized by  
the relativistic plasma is cool with the thermal velocity 
$u_{\rm crab}\sim10$~km s$^{-1}$.
The SN~Ia material in the adiabatic reverse shock  
is hot with the thermal velocity $u_{\rm sn}\sim10^4$~km s$^{-1}$.
The density contrast in the Crab is therefore by factor 
$(u_{\rm sn}/u_{\rm crab})^2\sim10^6$ larger. 

The behavior of RT spike may be affected by the KHI.
For a cylinder spike of the radius $a$ moving with the velocity $v$
along its own axis the perturbation growth time 
for the most destructive wave number $k\sim1/a$ 
is $t_{\rm KH}\sim(a/v)\chi^{1/2}$, where $\chi$ is the density ratio of  
SN and bubble material. 
At the SN radius $R=20$ pc the contrast $\chi\sim10^5$. The distance at 
which the 
most dangerous mode grows is then $\sim vt_{\rm KH}\sim a\sqrt{\chi}\sim 300a$. 
We do not aware of any multi-dimensional hydrodynamic simulations 
of a dense cloud moving in a rarefied relativistic fluid. A close analogue
is the two-dimensional hydrodynamic simulations 
of a dense cloud moving in the 
post-shock intercloud rarefied gas \citep{KMC94}. 
These simulations 
show that the cloud life time with respect to the fragmentation and 
fragment deceleration is order $\sim4(a/v)\chi^{1/2}$, roughly four times 
larger than the Kelvin-Helmholtz time.
Adopting this characteristic time one finds that the distance at which 
the spike will be destroyed and decelerated 
is $\sim 4vt_{KH}\sim 4a\chi^{1/2}\sim10^3a$.
Assuming $a/R\sim 10^{-2}$, comparable with the fingers in case of Crab nebula 
\citep{Hester96}, one finds that RT spike can travel 
$\sim10R\lesssim0.3$ kpc, rather small distance compared to the bubble radius.

On the other hand, \citet{Nu82} argues that 
the increase of the width of the boundary 
layer due to the KHI quenches the very instability. 
As a result the mass loss is suppressed and eventually is defined 
by the momentum transfer [cf. equations (\ref{eq-dotm}) and (\ref{eq-dotma})]. 
In this case 
the problem of mass stripping due to KHI is reduced to the problem of 
a deceleration of RT spikes which is analysed in the next section.

The longitudinal magnetic field also can suppress the KHI. Let $\rho_1$ and 
$\rho_2$ be the density of rarefied  and dense fluid. 
According to the criterion of the KHI in the presence of magnetic field 
\cite{Chan61} the condition that the magnetic field turns off
the KHI is 
\begin{equation}
B>(2\pi\rho_1)^{1/2}v=(6\pi p)^{1/2}\left(\frac{v}{c}\right)
\approx1.4\times10^{-6}~~\mbox{G}\,,
\end{equation}
where $p=10^{-10}$ erg cm$^{-3}$, $c$ is the light speed, and 
$v=10^9$ cm s$^{-1}$ are used.  
The required magnetic field $B>1.4\times10^{-6}$~G is within the range 
of field strength in the relativistic bubble, which can be as large as 
several $10^{-5}$ G. The magnetic stabilization of RT spikes against KHI thus 
seems quite plausible. 
Hereafter we address the deceleration of RT spike assuming its stability.

\subsubsection{Drag in collisionless case}

A typical RT spike is assumed to be a cylinder with the mass $m$, 
radius $a$, and length $b$, which are assumed to remain constant. 
The RT spike pesumably moves along the axis in the relativistic plasma 
dominated by the relativistic 
particles. With the giroradius $r_g\sim3\times10^{11}$ cm and the 
RT spike radius $a\sim10^{-2}r_{sn}\sim10^{18}$ cm only a motion along 
the regular magnetic field can be collisionless.   
The mean free path for relativistic protons propagating 
along the field can be constrained by scattering on the perpendicular 
component of a random field. 
The resulting mean free path along the mean field $B$ is 
$\lambda_{\parallel}\sim r_g(B/\delta B)^2$ \citep{SMP07}, 
where $\delta B$ is the amplitude of a random field with resonance wave number 
$k_{\rm res}r_g\sim1$.
Following \citet{SMP07} we assume the power law spectrum of 
random field energy density $W(k)\propto k^{-s}$ with $s=1.67$, 
between maximal length scale $k_{min}\sim 1/R_b$ and 
and minimal scale $k_{max}\sim 1/r_g$. The 
integrated energy density is normalized according to 
suggestion by \citet{SMP07}: $W=B^2/(8\pi)$. With 
these prerequisites
one gets $(\delta B/B)^2\sim(k_{min}/k_{\rm res})^{s-1}\sim10^{-7}$ and 
$\lambda_{\parallel}\sim 10^7r_g\sim3\times10^{18}$ cm, i.e., 
$\lambda_{\parallel}\gtrsim r_g$.
The situation thus is about collisionless, although uncertainties of 
the relevant parameters do not preclude collisional regime as well. 
One needs therefore to consider both collisionless and collisional cases. 

For $b\gg a$ the moment exchange occurs primarily 
via the cosmic ray collisions with 
the lateral surface of the spike. The momentum transferred to the colliding 
particle with the energy $E$ assuming diffusive reflection, is $\sim(E/c^2)v$. 
For the particle flux on the unit of surface area $(1/4)nc$,
where $n$ is the cosmic ray concentration, the moment transferred per second to 
all the striking protons (drag force) is 
$F_d=(1/2)\pi ab\epsilon(v/c)$, where $\epsilon=3p$ is 
the energy density of relativistic particles. Note that the 
drag force could be derived also
using average energy gain of a relativistic particle per collision with the 
cloud $E(v/c)^2$ 
\citep{F49}. Indeed, for a spherical cloud the total energy loss 
per second in that case is
$\pi a^2 ncE(v/c)^2=\pi a^2 \epsilon v^2/c$. This implies the drag force 
$\pi a^2\epsilon(v/c)$, which coincides with the above expression for 
$F_d$ within the geometrical factor. 

The equation of motion of the spike in the collisionless case with the above 
value of $F_d$ then reads 
\begin{equation}
m\frac{dv}{dt}=-1.5\pi abp\frac{v}{c}\,. 
\label{eq-eqmot}
\end{equation}
We neglect here the head-on collisions which would contribute the 
term of the order $\sim a/b\ll1$ in the right hand side. 
Note, the transition from spike to the spherical blob corresponds to $b=2a$
in the drag force expression. 
The characteristic time of the spike deceleration is thus
\begin{equation}
t_d\sim\frac{2}{3}\frac{mc}{\pi abp}\,.
\label{eq-tdrag}
\end{equation}
The ratio $m/\pi a^2$ can be expressed via the surface density of the SN shell 
at the stopping radius 
as $m/\pi a^2=\eta M/4\pi r_{sn}^2$, where the parameter $\eta\gg1$ 
because the spike is formed by a shell patch with the radius  $\gg a$. 
The deceleration distance for the spike is then
\begin{equation}
l_d=vt_d\sim\frac{10}{9}\eta r_{sn}\frac{c}{v}\frac{a}{b}=
1.2\eta\frac{a}{b}~~\mbox{kpc}\,,
\label{eq-rbdec}
\end{equation}
where we make use of equation (\ref{eq-rstop}) and adopt $v=10^9$ cm s$^{-1}$ 
and $r_{sn}=36$ pc. 
For $\eta\sim10$ and $a/b\sim0.1$ the spike can travel $\sim1$ kpc 
before it gets completely decelerated. In collisionless case the RT
fragments are decelerated efficiently inside the relativistic bubble.
It should be stressed, however, that the collisionless regime can be 
realized only in the case of the motion along the magnetic field 
so, only a small fraction of RT spikes experiences this type 
of a deceleration.

\subsubsection{Drag in collisional case}

If the mean free path for cosmic ray protons is small $\lambda\ll a$,
 one expects that the drag force should be proportional to $v^2$. 
The general condition for that is the large Reynolds number Re$>10^2$. 
To estimate Re we adopt the spike radius 
$a\sim10^{-2}r_{sn}\sim10^{18}$ cm, the spike velocity $v=10^9$ cm, and 
$\lambda=r_g$ as the mean free path for relativistic protons.
Assuming $B=10^{-5}$ G, i.e.,  $r_g\sim3\times10^{11}$ cm one 
gets Re$=3av/cr_g\sim 3\times10^5$. The condition Re$>10^2$ thus is 
fulfilled for $\lambda<3\times10^4r_g\sim 10^{15}$ cm, which is 
rather soft requirement. 

In the collisional approximation the drag force is $F_d=3\pi a^2p(v/c)^2$. 
Following the recipe of the previous section one obtains the 
spike deceleration distance 
\begin{equation}
l_d\approx\frac{5}{9}\eta r_{sn}\left(\frac{c}{v}\right)^2\,.
\end{equation}
For $v=10^9$ cm s$^{-1}$ and $r_{sn}=36$ pc one obtains $l_d\sim 17\eta$ kpc, 
rather large value that exceeds bubble radius (5 kpc) even 
for modest value of $\eta\sim 1$. The ram pressure drag thus 
almost does not decelerate RT spikes inside the relativistic bubble. 

\subsubsection{Alfven wave drag}

The Alfven wave drag can operate for the 
large conducting body $a>(v/c)r_g\sim10^{10}$ cm,  
which is easily met for RT spikes.
Using the expression for the power radiated in the form of Alfven waves 
\cite{Dre65}) one can write the Alfven drag force acting 
on the plasma spike with the radius $a$ and length $b$ as 
\begin{equation}
F_d=\frac{B^2}{2\pi}\frac{v}{v_{\rm A}}ab\,,
\end{equation}
where $v_{\rm A}=B/\sqrt{4\pi\rho_a}$ is the Alfven velocity. Note, 
it is only the 
lateral surface area ($\approx2ab$), that matters. Following arguments of the 
previous sections one gets the spike deceleration distance 
\begin{equation}
l_d\approx5\eta r_{sn}\left(\frac{\pi}{3}\right)^{3/2}\frac{\sqrt p}{B}
\frac{a}{b}
\frac{c}{v}=6\eta\frac{a}{b}B_5^{-1}p_{10}^{1/6}~~\mbox{kpc}\,.
\end{equation}
For fiducial model $p=10^{-10}$ erg cm$^{-3}$, $B=10^{-5}$ G, 
$v=10^9$ cm s$^{-1}$ one  
obtains $l_d\approx6\eta(a/b)$ kpc. 
This result shows that in case $a=b$ the blob can 
travel the distance exceeding the radius of the relativistic bubble even for 
moderate values of $\eta>2$. For a long spike ($b\gg a$) deceleration 
is by factor $b/a$ 
stronger and the deceleration distance is accordingly shorter. 
For $b/a\sim10$ and 
$\eta\sim 10$ the RT spike can travel $\sim6$ kpc, a distance 
comparable with the adopted bubble radius $R_b=5$ kpc. 
It takes roughly $R_b/v\sim 5\times10^5$ yr 
for the spike to reach the the bubble boundary assuming the spike average 
velocity of $10^4$ km s$^{-1}$. 

The effect of the Alfven wave drag is determined by the magnetic field. 
If the field is weak, $B=3\times10^{-6}$ G, the deceleration distance 
is $\sim18$ kpc, 
substantially larger than the bubble radius. On the other hand, for 
$B>10^{-5}$ G the deceleration distance is smaller than 
the bubble radius and significan amount of RT fragments will remain 
in the relativistic bubble. 
The escape probability for RT spike in case of $B=10^{-5}$ G
can be estimated adopting mean free path $\lambda=vt_d=6$ kpc,
i.e., $\tau=R_b/\lambda=5/6$. Equation (\ref{eq-oster}) gives 
in this case the escape probability $p_{\rm esc}\approx0.6$. 

\subsubsection{Spike deceleration by wind material}

The average number of wind shells along the 
average distance to the bubble boundary for the fiducial model is 
$\sim0.08$ (Section \ref{sec-windyn}) 
which means that for SN fragments the probability to collide 
with a wind shell is low, $\sim0.06$. A question arises, 
what happens, anyway, if the collision takes place.  

The deceleration is determined by the ratio of column densities ($\mu$) of the 
projectile (RT spike) and target (wind shell). For the wind shell 
\begin{equation}
\mu_w=\frac{m_w}{\pi r_w^2}=1.4\times10^{-4}~~\mbox{g cm}^{-2}\,,
\end{equation}
where $r_w=1.5\times10^{18}$ cm and $m_w=0.5~M_{\odot}$ are used. The column 
density of the 
spike $\mu_s=\eta M/(4\pi r_{sn}^2)\sim2\times10^{-8}\eta$ g cm$^{-2}\ll\mu_w$. 
This comparison shows that 
the spike will be fully decelerated in a single collision with a wind shell. 
We conclude therefore that for the fiducial model
there is non-negligible probability, $\sim0.06$, that the spike will be 
decelerated in the bubble via collision with the wind shell. 

\begin{table}[t]
\caption[]{Parameters of fiducial model}
\label{tab-numbers}
\centering
\begin{tabular}{ l l c }
\hline\hline
\noalign{\smallskip}
Parameter & Description & Value \\
\noalign{\smallskip}
\hline
\noalign{\smallskip}
$R_b$     & Bubble radius & 5 kpc \\
$t_b$     & Age   & $3\times10^7$ yr  \\
$p $      & Pressure   & $10^{-10}$ erg cm$^{-3}$  \\
$B$       & Magnetic field  & $10^{-5}$ G \\
$n $      & Number density of ICM &  0.02 cm$^{-3}$ \\
$M_s$     &  Stellar mass in the bubble  & $9\times10^{10}~M_{\odot}$  \\
$\dot{N}$ & Stellar death rate   &  0.83 yr$^{-1}$  \\ 
$\dot{N}_{sn}$  & SN~Ia rate   &  0.0066 yr$^{-1}$ \\
$r_w$     & Wind stopping radius    &  0.5 pc   \\ 
$r_{sn}$      &  SN stopping radius   &  36 pc \\ 
\noalign{\smallskip}
\hline
\end{tabular}
\end{table}

\section{Discussion}

The aim of this paper has been to get an idea on what happens to the 
matter ejected
by the stellar population of BCG embedded into a bubble of
relativistic plasma. We have found that the expansion of a wind envelope
lost by a star is stopped by the pressure of the relativistic fluid
when the radius attains $\sim0.5$ pc. Because of the small size of the 
wind shell its bulk motion is not decelerated neither by the ram
pressure, nor by the Alfven wave drag, so the shell escapes the bubble
together with the parent star.  The SN~Ia exploding in the
relativistic fluid expands up to much larger radius $\sim30-40$ pc
and its bulk motion can be efficiently decelerated by the Alfven
wave drag. Unless a RT instability operates, the SN material would not
escape the relativistic bubble, but will instead be advected by the
buoyantly rising relativistic fluid. 

The RT instability can strongly
modify the behavior of the ejecta.  In the framework of our fiducial
model we find that significant fraction of SN fragments escapes
the relativistic bubble. In
our analysis of this scenario we rely on the fiducial model 
parameters outlined in Table 1.
A gas lump crossing the bubble boundary and entering
the ICM is decelerated after sweeping the ICM mass comparable
with its own mass. The escaping wind shell gets
decelerated in the thermal plasma of ICM at the length of $l_w\sim
1$ kpc (Section \ref{sec-windyn}). 
The wind deposits $\sim1.4\times10^7~M_{\odot}$ in this layer. 
This value should be compared with the mass of the ICM gas, which is 
already there. For $R_b=5$ kpc and hydrogen number density $n=0.02$ cm$^{-3}$ 
the mass of the ICM in the boundary layer with the thickness 
$l_w=1$ kpc is $7\times10^7~M_{\odot}$, i.e.,  
factor $\sim5$ larger than the mass deposited by the wind shells.
The deposited wind mass scales with the bubble radius as $\propto R_b^4$, 
whereas the ICM mass in the layer is $\propto R_b^2$.
We thus conclude that for the fiducial model
the wind material escaping the relativistic bubble does 
not change significantly the density of the surrounding ICM; moreover 
the effect even smaller in case of M~87 in which bubble radius 
$R_b\sim1.4$ kpc.
The effect of the energy deposition by the escaping wind shells 
is also negligible because the bulk velocities of the wind shells 
is subsonic and the deposited mass is low. The chemical composition of ICM
is not affected by the escaping wind shells either. 

Unlike the wind shells, fragments of SN ejecta escaping the relativistic 
bubble may have a profound effect on the enrichment of ICM 
by iron peak elements. The resulting abundance is determined by 
the width of the mixing layer. Generally, one should consider 
time-dependent model of the formation of mixing layer. However, we 
assume that the mixing layer in the fiducial model is formed 
by the cumulative effect of all SNe.    
Emploing the momentum conservation arguments a deceleration distance for 
the RT spike entering the ICM with the 
velocity $v_i$ and decelerated down to $v_f$ turns out to be 
\begin{equation}
l_{\rm sn}\sim\eta\ln(v_i/v_f)\frac{M}{4\pi r_{sn}^2\rho}
\approx0.5\eta~~\mbox{pc}\,,
\end{equation}
where we used $v_i/v_f=20$, $n=0.02$ cm$^{-3}$, $M=1.4~M_{\odot}$, 
and $r_{sn}=36$ pc.  For $\eta=10$ one gets the 
deceleration length $l_{\rm sn}\sim 5$ pc. At first glance 
the mixing layer could be identified with the deceleration layer. 
This layer 
contains $M_{icm}\sim7\times10^5~M_{\odot}$ of the ICM gas. 
The mass produced by SN~Ia during the life time of the bubble is
$M_{sn}\sim3\times10^5~M_{\odot}$. If most of the SN mass escapes the 
bubble, the amount of escaping iron in the mixing layer 
turns out to be $\sim10^5~M_\odot$ which corresponds to the 
iron abundance $\sim80\times$(solar). 

On the other hand, the mixing layer could be broader 
because the total volume of shocked SN fragments in pressure 
equilibrium substantially exceeds the volume of the deceleration layer. 
Simple estimate based upon the pressure equilibrium suggests the total 
volume occupied by shocked SN ejecta to be $V_t=N_{sn}E/p$ which implies 
the layer width $\sim300$ pc. If this is identified with 
the width of the mixing layer than the iron abundance will be 
factor two larger compared to solar abundance of pre-existing ICM.
The increase of the iron abundance by factor two changes the 0.6-2 keV
emissivity\footnote{ 0.6-2 keV is the energy range where present day
  grazing incidence X-ray telescopes are most sensitive} by factors of
1.8, 1.5 and 1.3 for the gas temperatures 1, 2 and 3 keV respectively. 

The latter estimates suggest complete mixing of injected iron with the ICM, 
which may not be the case. The point is that the deceleration of SN 
fragments in the ICM results in the strong
heating of the ejecta material up to temperature corresponding to its 
kinetic energy. Most of the iron therefore ends up
in the high entropy/low density gas with the very little X-ray emission. 
The observational outcome thus critically depends on the mixing degree 
between hot SN gas and the relatively cool ICM. 

The above picture of mixing layer is very crude and one 
cannot rule out a possibility that the expanding
relativistic bubble continuously catches up with the mixing layer 
outer boundary so that most of the decelerated SN material injected 
in the ICM eventually turns out engulfed by the relativistic fluid. 
Qualitatively this scenario then gets 
similar to the case when SN~Ia ejecta bulk motion is decelerated well
inside the bubble. In this case no strong X-ray emission is expected
from fully expanded shells (because of the very low gas density). The
fate of the iron generated by SN~Ia in this scenario solely depends on
the evolution of relativistic plasma, which can escape the central
region of the galaxy, as suggested by observations of e.g. M87
\citep{Chur01, For05}.

\section{Conclusions}

We consider the outcome of the mass ejection by stars via winds and
supernovae inside a bubble of relativistic plasma inflated by an AGN
in the core of BCG. Wind shells are likely to escape the
relativistic bubble and deposit their mass in the ICM within 
$\sim1$ kpc from the bubble boundary. SN~Ia exploded inside the
bubble is efficiently decelerated owing to the pressure of the
relativistic fluid.  If the SN shells remain spherical until the
expansion of the envelope stops and do not fragment, then they would
not escape the bubble. In this case the iron produced by SN~Ia is
advected by the relativistic plasma and may leave the central region
of the BCG together with buoyantly moving bubbles.

As a possibility we consider a scenario in which the RT instability of
the SN envelope at the deceleration phase breaks the shell into
multitude of RT spikes. The analysis of the deceleration of RT spikes
in the relativistic fluid shows that the SN fragments are able to
escape the bubble. The fragments are decelerated in the ICM in a close
vicinity of the bubble boundary thus producing Fe-rich layer.  In the
optimistic scenario this Fe-rich layer can enhance X-ray emission
around bubbles of relativistic plasma, producing bright rims around
bubbles.

\section{Acknowledgements}

We are grateful to Nail Inogamov and Sergey Sazonov for useful 
discussions, and to Ewald M\"uller for sharing hydrocode.
NC thanks Wolfgang Hillebrandt for the invitation to MPA. 
The work was partly supported by the Division of Physical Sciences 
of the RAS (the program ``Extended objects in the Universe'', OFN-16) 
and the project NSH-5069.2010.2.



\begin{thebibliography}{}
\bibitem[\protect\citeauthoryear{Birzan et al.}{2004}]{ma99}
 Matthews J.M., Kurtz D.W., Martinez P., 1999, ApJ, 511, 422
\bibitem[\protect\citeauthoryear{Birzan et al.}{2004}]{Bir04}
 Birzan, L., Rafferty, D. A., McNamara, B. R., Wise, M. W., Nulsen, P. E. J.
   2004, ApJ, 607, 800
\bibitem[\protect\citeauthoryear{B\"{o}hringer et al.}{1993}]{Boer93}
 B\"{o}hringer, H., Voges, W., Fabian, A. C., Edge, A. C.,  Neumann, D. M.
 1993, MNRAS, 264, L25	
\bibitem[\protect\citeauthoryear{Chandrasekhar}{1961}]{Chan61}
 Chandrasekhar, S. 1961.  Hydrodynamic and hydromagnetic stability. 
 Oxford University Press 
\bibitem[\protect\citeauthoryear{Churazov et al.}{2000}]{Chur00}
 Churazov, E., Forman, W., Jones, C., B\"{o}hringer, H.
 2000, A\&A, 356, 788
\bibitem[\protect\citeauthoryear{Churazov et al.}{2001}]{Chur01}
 Churazov, E., Br\"{u}ggen, M., Kaiser, C. R., B\"{o}hringer, H., 
 Forman, W. 2001, ApJ, 554, 261
\bibitem[\protect\citeauthoryear{Drell et al.}{1965}]{Dre65}
 Drell, S. D. Foley, S. D., Ruderman, M. A. 1965, JGR, 70, 3131
\bibitem[\protect\citeauthoryear{Fabian et al.}{2003}]{Fab03}  
 Fabian, A. C., Sanders, J. S., Crawford, C. S., Conselice, C. J., 
 Gallagher, J. S.,  Wyse, R. F. G. 2003, MNRAS, 344, 48
\bibitem[\protect\citeauthoryear{Fermi}{1949}]{F49}
 Fermi, E. 1949, PhRv, 75, 1169
\bibitem[\protect\citeauthoryear{Forman et al.}{2005}]{For05}
 Forman, W., Nulsen, P., Heinz, S., Owen, F., Eilek, J., Vikhlinin, A.,
 Markevitch, M., Kraft, R., Churazov, E., Jones, C.
 2005, ApJ, 635, 894
\bibitem[\protect\citeauthoryear{Forman et al.}{2007}]{For07}
 Forman, W., Jones, C., Churazov, E., Markevitch, M., Nulsen, P., 
 Vikhlinin, A., Begelman, M., B\"{o}hringer, H., Eilek, J., 
 Heinz, S., Kraft, R., Owen, F., Pahre, M. 2007, ApJ, 665, 1057
\bibitem[\protect\citeauthoryear{Gedalin}{1993}]{Ged93}
Gedalin, M. 1993, Phys. Rev. E, 43, 4354
\bibitem[\protect\citeauthoryear{Gull \& Northover}{1973}]{GN73}
 Gull, S. F., Northover, K. J. E. 1973, Nature, 244, 80
\bibitem[\protect\citeauthoryear{Hester et al.}{1996}]{Hester96}
 Hester, J. J., Stone, J. M., Scowen, P. A., et al. 1996, ApJ, 456, 225
\bibitem[\protect\citeauthoryear{Huang \& Sarazin}{1998}]{HS98} 
 Huang, Z. ,  Sarazin, C. 1998, ApJ, 496, 728
\bibitem[\protect\citeauthoryear{Johnson et al.}{2009}]{John09} 
 Johnson, R., Chakrabarty, D., O'Sullivan, E., Raychaudhury, S.
 2009, ApJ, 706, 980
\bibitem[\protect\citeauthoryear{King}{1966}]{King66}
 King, A. R.  1966, AJ, 71, 64
\bibitem[\protect\citeauthoryear{Klein et al.}{1994}]{KMC94} 
 Klein, R. I., McKee, C. F., Colella, P. 1994, ApJ, 420, 213
\bibitem[\protect\citeauthoryear{Lou et al.}{2010}]{LZ10} 
 Lou, Y.- Q., Zhai, X. 2010, MNRAS, 408, 436
\bibitem[\protect\citeauthoryear{Mathews \& Brighenti}{2003}]{MB03}
 Mathews, W. G., Brighenti, F. 	2003, ApJ, 599, 992
\bibitem[\protect\citeauthoryear{McNamara et al.}{2000}]{McN00} 
 McNamara, B. R. et al. 2000, ApJL, 534, L135
\bibitem[\protect\citeauthoryear{Narayan \& Medvedev}{2001}]{Nar01} 
 Narayan, R., Medvedev, M. V. 2001, ApJ, 562, L129	
\bibitem[\protect\citeauthoryear{Nulsen}{1982}]{Nu82} 
 Nulsen, P. E. J. 1982, MNRAS, 198, 1007
\bibitem[\protect\citeauthoryear{Osterbrock}{1989}]{Ost89} 
 Osterbrock, D. E. 1989. Astrophysics of gaseous nebulae and 
 active galactic nuclei. University science books, p.385
\bibitem[\protect\citeauthoryear{Pritchet et al.}{2008}]{Prit08}  
 Pritchet, C. J., Howell, D. A.,  Sullivan, M. 2008, ApJL, 683, L25
\bibitem[\protect\citeauthoryear{Richer et al.}{2008}]{Rich08}  
 Richer, M. G., Lopez, J. A., Pereyra, M., Riesgo, H., Garcia-Diaz, M. T.,
 \& Baez, S.-H. 2008, ApJ, 689, 203 
\bibitem[\protect\citeauthoryear{Salaris al.}{2009}]{Sala09}   
 Salaris, M., Serenelli, A., Weiss, A.,  Miller Bertolami, M.
 2009, ApJ, 692. 1013
\bibitem[\protect\citeauthoryear{Sanders \& Fabian}{2007}]{Sand07}    
 Sanders J.~S., Fabian A.~C., Heating  versus Cooling in Galaxies and 
 Clusters of Galaxies, ESO Astrophysics Symposia, 
 Springer-Verlag Berlin Heidelberg, 2007, 74
\bibitem[\protect\citeauthoryear{Schaller et al.}{1992}]{Scha92} 	
 Schaller, G., Schaerer, D., Meynet, G., Maeder, A.
 1992, A\&AS, 96, 269
\bibitem[\protect\citeauthoryear{Stage et al.}{2006}]{Stage06} 
Stage, M. D., Allen, G. E., Houck, J. C., Davis, J.E. 2006, 
 Nature Phys., 2, 614 
\bibitem[\protect\citeauthoryear{Steffen et al.}{1998}]{Stef98} 
  Steffen, M., Szczerba, R.,  Schoenberner, D. 1998, A\&A, 337, 149
\bibitem[\protect\citeauthoryear{Strong et al.}{2007}]{SMP07} 
 Strong, A. W., Moskalenko, I. V., Ptuskin, V. S. 2007, ARNPS, 57, 285 
\bibitem[\protect\citeauthoryear{Vassiliadis \& Wood}{1993}]{VW93}
 Vassiliadis, E.,  Wood, P. R. 1993, ApJ, 413, 641
\bibitem[\protect\citeauthoryear{Wu \& Tremaine}{2006}]{WT06} 
 Wu, X., Tremaine, S. 2006, ApJ, 643, 210
  
\end{thebibliography}
\end{document}